
\input harvmac
\newcount\figno
\figno=0
\def\fig#1#2#3{
\par\begingroup\parindent=0pt\leftskip=1cm\rightskip=1cm\parindent=0pt
\baselineskip=11pt
\global\advance\figno by 1
\midinsert
\epsfxsize=#3
\centerline{\epsfbox{#2}}
\vskip 12pt
{\bf Fig. \the\figno:} #1\par
\endinsert\endgroup\par
}
\def\figlabel#1{\xdef#1{\the\figno}}
\def\encadremath#1{\vbox{\hrule\hbox{\vrule\kern8pt\vbox{\kern8pt
\hbox{$\displaystyle #1$}\kern8pt}
\kern8pt\vrule}\hrule}}

\overfullrule=0pt

%

%

\font\zfont = cmss10 

\def\bigone{\hbox{1\kern -.23em {\rm l}}}
\def\ZZ{\hbox{\zfont Z\kern-.4emZ}}

\Title{hep-th/9506101, IASSNS-HEP-95-51}
{\vbox{\centerline{STRONG COUPLING}\smallskip
\centerline{AND THE COSMOLOGICAL CONSTANT}}}
\smallskip
\centerline{Edward Witten}
\smallskip
\centerline{\it School of Natural Sciences, Institute for Advanced Study}
\centerline{\it Olden Lane, Princeton, NJ 08540, USA}\bigskip
\baselineskip 18pt

\medskip

\noindent
The vanishing of the cosmological constant
and the absence of a massless dilaton might be explained by a duality
between a supersymmetric string vacuum in three dimensions
and a non-supersymmetric string vacuum in four dimensions.

\Date{June, 1995}

One of the central mysteries in physics is why the cosmological
constant vanishes without observable supersymmetric bose-fermi degeneracy.
Recently \ref\witten{E. Witten, ``Is Supersymmetry Really
Broken?'' Int. J. Modern Phys. {\bf A10} (1995) 1247, hepth-9409111.},
it was pointed out that in $2+1$ dimensions the problem can
be more or less addressed as follows.
Unbroken supersymmetry can lead
to natural vanishing of the cosmological constant in $2+1$ dimensions
(or any dimension), under certain conditions.
On the other hand, in $2+1$ dimensions, unbroken supersymmetry
does {\it not} lead to bose-fermi degeneracy.
The reason for the latter statement is that in $2+1$ dimensions,
any non-zero mass creates a conical structure at infinity, which
prevents one from defining global conserved supercharges.
(In some cases with $N>1$ supersymmetry in three dimensions
one finds \ref\strom{K. Becker, M. Becker, and
A. Strominger, ``Three-Dimensional Supergravity And The Cosmological
Constant,'' hepth-9502107.} that some global supercharges can be
defined for certain soliton states, but not enough to imply
bose-fermi degeneracy.)

This scenario has two fairly obvious flaws:

(1) It uses facts that are very special to three dimensions.

(2) Even in three dimensions, the mass splittings produced this
way are suppressed by powers of Newton's constant if (as in the real
four-dimensional world) gravity is weakly coupled and one considers
particles with mass far below the Planck mass.

In this paper, I will make an attempt to improve the scenario,
using a recent observation that a $d$-dimensional theory can become
$d+1$-dimensional in a strong coupling limit \ref\uwitten{E. Witten,
``String Theory Dynamics In Various Dimensions,'' hepth-9503124, to appear
in Nucl. Phys. B.}.  As a bonus, the scenario will also suggest a possible
way to eliminate the massless dilaton from the string theory spectrum
(avoiding contradictions with tests of general relativity)
without the gymnastics involved in trying to generate a non-trivial,
stable dilaton potential.  The scenario is based on optimistic extrapolations
of recent experience about string theory duality, and is unlikely to be
precisely correct in its present form.

Before launching into the discussion, let us recall
the strong coupling
behavior argued in \uwitten\ for Type IIA superstrings in ten dimensions
and for the heterotic string toroidally compactified
to five dimensions.  In each of those cases,
when the string coupling
$\lambda$ is large one gets an effective description of the $d$-dimensional
theory as a $d+1$-dimensional theory on ${\bf R}^d\times {\bf S}^1$,
with the radius of the ${\bf S}^1$ being a positive power of $\lambda$.
The limit $\lambda\to \infty$, if it exists, should therefore be a Poincar\'e
invariant $d+1$-dimensional theory.  For five-dimensional heterotic
strings, the limit exists and is the Type IIB theory in six dimensions
(which incidentally is chiral even though it arises as a strong coupling
limit of a non-chiral theory in five dimensions).
For the Type IIA theory in ten dimensions, we do not know if
the large $\lambda$ limit exists; if it does it would be a new
quantum theory -- perhaps related to supermembranes --
with eleven-dimensional supergravity as its low energy limit.
In any case, whenever we manage to take
the large $\lambda$ limit, the original
dilaton disappears -- having been fixed at $+\infty$
to get a Poincar\'e invariant theory in $d+1$ dimensions.  For
the five-dimensional heterotic string, but not for the ten-dimensional
Type IIA theory, there is an additional scalar that behaves as a dilaton
in the $d+1$-dimensional world; that is perhaps the main reason that
the five-dimensional case is better understood.

Now let us return to our problem.
We begin in three dimensions with a string vacuum with unbroken
supersymmetry, a dilaton $\phi$, and a string coupling constant
$\lambda= e^\phi$.
We suppose that the theory has a sufficient amount of supersymmetry
or suitable discrete symmetries so that unbroken supersymmetry is stable.
(For instance, this will be the case in an $N=1$ theory with a discrete
CP symmetry -- likely in the scenario below because it comes from
CPT in four dimensions -- if all massless chiral superfields are CP even.
That is because the superpotential is CP odd for $N=1$ theories in
$2+1$ dimensions.)

For small $\lambda$, string perturbation theory
is good and the cosmological constant is zero because of unbroken
supersymmetry.  Bosons and fermions are not degenerate, but their
mass splittings are very tiny.  As $\lambda$ increases, the bose-fermi
splittings become large, but the cosmological constant remains zero
because of supersymmetry.

Now what happens for $\lambda\to\infty$?  We assume that
as in the five- and ten-dimensional examples cited above, the large
$\lambda$ limit of the three-dimensional theory is a Poincar\'e invariant
theory in four dimensions.  The cosmological constant is zero in this
four-dimensional theory since it is in fact identically zero for all
$\lambda$.  The bose-fermi mass splittings were non-zero for all finite
$\lambda$ so hopefully they remain non-zero for
$\lambda\to\infty$.  (This claim is not as innocent as it may sound
since on ${\bf R}^3\times {\bf S}^1$ there appear to be light states
as the radius of the ${\bf S}^1$ goes to infinity.
One of the key points is really to find a situation in which the
claim holds.)
Finally, the three-dimensional dilaton disappears when we take
$\lambda\to\infty$, as was explained above, so if (as in the Type IIA
theory in ten dimensions) there is no additional field with
the right properties to take its place, the problem of the dilaton is solved.

If this scenario is valid, the three-dimensional description of nature
has the virtue of explaining the vanishing of the cosmological
constant, but (since the real world is infinitely strongly coupled
from the three-dimensional point of view) it may have little value for
explaining anything else in nature.  To get
some predictive power concerning the rest of physics, one would need
a dual description of the real world, or perhaps several dual descriptions,
that might explain some phenomena while perhaps leaving the
cosmological constant as a mystery.

It has been suggested (by C. Vafa) that the three-dimensional
theory entering the above scenario
may be the one required  in proposals for a holographic
theory of the universe \ref\susskind{L. Susskind,
``The World As A Hologram,'' hepth-9409089.}.

\bigskip

I would like to thank N. Seiberg, L. Susskind, and C. Vafa for discussions,
and to acknowledge the hospitality of the International Center
for Theoretical Physics in Trieste.

\listrefs
\end